\documentclass[letterpaper, 10 pt, conference]{ieeeconf}  

\IEEEoverridecommandlockouts                              

\usepackage{graphicx} 
\usepackage{subcaption}
\usepackage{amsmath} 
\usepackage{amssymb}  
\usepackage[T1]{fontenc}
\usepackage{amsfonts}
\usepackage{tikz}
\usepackage{url}
\newcommand{\real}{\mathbb{R}}
\title{\LARGE \bf
Algorithmic collusion in a two-sided market: A rideshare example*
}

\author{Pravesh Koirala$^{1}$ and Forrest Laine$^{1}$
\thanks{*This work was not supported by any organization}
\thanks{$^{1}$Department of Computer Science,
        Vanderbilt University, Nashville, TN, USA
        {\tt\small pravesh.koirala@vanderbilt.edu, forrest.laine@vanderbilt.edu}}%
}

\begin{document}

\maketitle
\thispagestyle{empty}
\pagestyle{empty}

\begin{abstract}
With dynamic pricing on the rise, firms are using sophisticated algorithms for price determination. These algorithms are often non-interpretable and there has been a recent interest in their seemingly emergent ability to tacitly collude with each other without any prior communication whatsoever. Most of the previous works investigate algorithmic collusion on simple reinforcement learning (RL) based algorithms operating on a basic market model. Instead, we explore the collusive tendencies of Proximal Policy Optimization (PPO), a state-of-the-art continuous state/action space RL algorithm, on a complex double-sided hierarchical market model of rideshare. For this purpose, we extend a mathematical program network (MPN) based rideshare model to a temporal multi origin-destination setting and use PPO to solve for a repeated duopoly game. Our results indicate that PPO can either converge to a competitive or a collusive equilibrium depending upon the underlying market characteristics, even when the hyper-parameters are held constant. 

\end{abstract}

\section{INTRODUCTION}

Dynamic pricing has seen a large uptrend ever since internet-enabled handheld devices became a commonplace. Smartphones, equipped with GPS and anytime connectivity, can continuously push location (and other relevant) statistics to centralized servers which can then estimate a localized demand pattern. Based on this demand information, sophisticated machine learning models can instantaneously produce an optimal price for any given spatio-temporal region. Industries like gasoline, ticketing, online retail, etc are known to use dynamic pricing \cite{garbarino2003dynamic}. Recently, platform economies like rideshares are using dynamic pricing to hike prices during period of high demands i.e. \textit{surge}, with the intent of ensuring high service availability to requests of priority. It is argued that in platform economies like rideshares, dynamic pricing produces better outcomes both in shorter timeframe by incentivizing the drivers to accept more rides, and in the longer timeframe by shaping driver's schedules to identified surge periods \cite{chendynamic}. Regardless, increasing number of industries seem to be gravitating towards dynamic pricing, including food giants like Wendys who recently withdrew plans for proposed surge pricing following a sharp backlash \cite{bbc2024}.

With dynamic pricing gaining traction, much has been invested in devising complex models that can accurately predict an optimum price given informative attributes of any spatio-temporal frame. However, most of the sophisticated machine learning models are effectively a black-box \cite{rudin2019stop}. They are known to learn non-trivial latent representation of the given dataset and often, these representations are not readily interpretable. A recent concern with these machine learning based pricing agents is the possibility of tacit collusion. Collusion refers to the tendency of rational agents to cooperate with each other with the intent of earning profits that would be higher than if they were competing i.e. supra-competitive profits. While in most of the economies around the world, a coordinated collusion among firms is heavily penalized in the spirit of promoting healthy competition and maintaining a free market economy, emergent collusion in pricing algorithms is a perplexing phenomenon from not only regulatory but also technological perspective. Perhaps of most importance is the fact that it's not entirely clear what causes a collusive behavior to emerge in disconnected algorithmic agents. Some works argue that certain hyperparameters give rise to collusion \cite{waltman2008q}, while others argue that superior algorithms cause this behavior \cite{brown2023competition}. Irrespective of these unknowns, evidence suggests that algorithmic collusion does exists, with a study claiming an increase of almost 28\% margins in the German retail gasoline market after adoption of the dynamic pricing by duopolies \cite{assad2024algorithmic}.

Much of the related literature have investigated simple algorithms like Q-learning in a simpler market setting like single-sided duopoly competition to argue about potential collusion, but markets are often complex (hierarchical, multi-sided, etc.) and industries are moving towards increasingly powerful methods to optimize their profits. It's not apparent, then, if observed collusive behavior in established literature also emerges in such non-trivial market regimes under modern algorithms. Also, to our knowledge, there is no prior work that studies algorithmic collusion as function of market responsiveness i.e. the rate at which a market adjusts to pricing changes. We believe that this is an important market attribute that warrants exploration as it signifies how \textit{aware} a market is of the underlying pricing algorithms.
\setcounter{footnote}{1}
In this paper, we study collusive behaviors in sophisticated pricing algorithms operating in a complex platform market i.e. rideshare \footnote{Used synonymously to ridehailing throughout this work as per established usage of the word}. We choose rideshare for this study as it is a flagship example of a hierarchical market with its two networks (passengers and drivers) and various inter/intra network externalities. We begin with a mathematical program network (MPN) \cite{laine2024mathematical} based model of rideshare first presented by Koirala and Laine \cite{koirala2023decreasing} and extend it to a dynamic multi origin-destination setting.  Since we want to investigate  a modern reinforcement learning algorithm for potential collusion, we use Proximal Policy Optimization (PPO) \cite{schulman2017proximal}, which operates in a continuous state/action space and is a much more realistic algorithm for pricing agents in today's context. We further extend our study by doing a comparison on the collusive behavior of the agents under different intrinsic market characteristics. In summary, our major contributions are as follows:
\begin{itemize}
    \item We investigate potential collusive behaviors in a non-trivial market with modern reinforcement learning algorithms.
    \item We compare the emergent collusive behavior under different market responsiveness.
\end{itemize}

To the best of our knowledge, these contributions are novel and add substantively to the literature of algorithmic pricing and collusion. The rest of the paper is organized as follows: In section \ref{sec:litreview} we present a brief overview of existing literature on algorithmic collusion. In section \ref{sec:model}, we describe our MPN based model and in section \ref{sec:setups}, we explain our experimental setup. Finally, in section \ref{sec:results}, we describe our obtained results and conclude our arguments in \ref{sec:conclusions}. 

\section{Literature Review}
\label{sec:litreview}
Huck, et al. \cite{huck2004through} first showed that simple learning agents using a trial-and-error approach can learn to collude without even having the knowledge of payoff function in a cournout duopoly. The seminal study by Waltman, et al. \cite{waltman2008q} showed, via simulation, that Q-learning agents can learn to collude with each other in Cournot oligopoly games and while they may not learn optimal collusive behavior, their joint profit is better than that of perfect competition. They also analytically argue that potential collusion are results of exploratory tendencies of algorithms. Salcedo \cite{salcedo2015pricing} showed that in a dynamic game where firms use a fixed pricing algorithm that can be revised over time and be decoded by competitors, collusion is not only possible but is inevitable. 

Perhaps the most defining work in algorithmic pricing is by Calvano, et al. \cite{calvano2020artificial}, where they argued that Q-learning agents in an oligopoly model learn to charge supracompetitive prices without any explicit communication with each other. The distinguishing feature of their work is that they find that collusions among agents are enforced via punishments in case of deviation. A counter argument is proposed by Miklos-Thal and Tucker \cite{miklos2019collusion} where they argue that algorithms with better demand forecasting can, in fact, lead to an increase in social welfare. Brown and MacKay \cite{brown2023competition} performed an study on the pricing behavior of firms with assymetric pricing technology. They showed that if a single firm adopts superior pricing technology, all firms are able to obtain higher prices and if all firms adopt automated high-frequency algorithms, collusive pricing emerges. Klein \cite{klein2021autonomous} presented a different pricing model based on Q-learning where agents, instead of deciding on prices simultaneously, take a turn-based approach. They found that even under this market regime, collusive pricing does emerge. Abada and Lambin \cite{abada2023artificial} extended results from \cite{waltman2008q, calvano2020artificial} to a multi-agent pricing game to buy and sell a storable good. They argued that seeming collusion may originate in imperfect exploration (where algorithms lock in rapidly to a state of exploitation) and not algorithmic sophistication. A work by Boer, et al. \cite{den2022artificial} questioned the results obtained by Calvano \cite{calvano2020artificial} and argued that simple Q-learning algorithms are not a cartel concern. Sanchez-cartas and Katsamakas \cite{sanchez2022artificial} compared Particle Swarm Optimization algorithm and Q-learning algorithm for logit, hotelling, and linear demand models and find that PSO sets a more competitive price as compared to Q-learning model.

As mentioned earlier, most of these works are based on simple models like Q-learning and PSO. Hettich \cite{hettich2021algorithmic} replicated the results of \cite{calvano2020artificial} using a Deep Q-Network based architecture and showed significantly fast convergence and, under certain state reformulation, collusive outcomes for wider oligopolies (i.e. with more participants). However, their model is still a discrete space model and the first continuous space model to investigate algorithmic collusion seems to be from Wang \cite{wang2022will} who used a natural policy gradient method to demonstrate a counter result i.e., they show that that their policy gradient based implementation converges to the competitive Nash equilibrium. They go on to suggest that algorithmic collusion might be a feature of the learning algorithm being used. Our study, like Wang \cite{wang2022will}, uses a continuous space reinforcement learning algorithm i.e. PPO. But unlike other existing works with focus on simple single-sided market duopolies, we model a complicated two-sided market like ridesharing that has inherent externalities. Furthermore, we primarily study the role of market responsiveness, i.e. the rate at which the market adjusts to pricing decisions, in the outcome of the equilibrium and potential collusion.

\section{Model}
\label{sec:model}
The ridesharing economy has ample moving parts. First, there are the strategic agents such as \textit{platforms} like Uber, Lyft, Didi, etc. who decide what to charge and how much to pay to, respectively, their \textit{passengers} and \textit{drivers}, who, in turn, decide which platform to use and serve. When choosing a platform to utilize, many of the passengers and drivers are typically not tied to a specific platform i.e. they can use any of the available ones depending upon the costs and the profits and, thus, are engaged in what's called a \textit{multi-homing} behavior adding further complexity to the model. With the drivers forming the supply network and the passengers forming the demand network, any model must also capture the interactions (externalities) within and across these two networks, i.e. the additional utility / cost derived by any participant of the network from the presence / absence of participants in the same (same-side) or the complementary (cross-side) network. Concretely, a passenger is better served when there are more drivers in a platform due to reduced wait-times and vice-versa, which shows that there is a positive cross-side externality between the supply and demand networks. However, a passenger (driver) is worse-off when there are more passengers (drivers) because of increased wait-times (decreased utilization) which is indicative of a negative same-side externality. A final modeling difficulty is due to the order of decision as there may be different equilibrium outcomes in the game depending upon who moves/commits to an action first.

In consideration of these challenges, we use the Mathematical Program Network (MPN) based ridesharing model introduced by Koirala and Laine \cite{koirala2023decreasing} (other game theoretical models of rideshare can be found in \cite{nikzad_thickness_2017, bernstein_competition_nodate, bryan_theory_2019, bai_can_2022}). This model essentially consists of a partially ordered directed graph of mathematical optimizers which captures both the negative same-side externalities and the positive cross-side externalities inherent in the ridesharing market. Furthermore, in this model, the rideshare game is represented as a sequential game of three stages with the platforms moving first, followed by the drivers, and then the passengers. This scheme highlights the importance of the drivers as their allocation decision conclusively shapes the equilibrium outcome of the entire game. Lastly, this model has clear interpretations under both single-homing and multi-homing regimes.

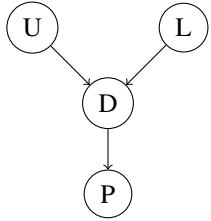
\begin{figure}[h]
    \centering
    \begin{tikzpicture}
        \node[draw, circle] (U) at (-1,1) {U};
    
        \node[draw, circle] (L) at (1,1) {L};
        \node[draw, circle] (D) at (0,0) {D};
        \node[draw, circle] (P) at (0,-1.2) {P};
        \draw[->] (U) -- (D);
        \draw[->] (L) -- (D);
        \draw[->] (D) -- (P);
    \end{tikzpicture}
    \caption{MPN based model for the ridesharing duopoly problem proposed by \cite{koirala2023decreasing}. Platforms U and L decide wages and rates simultaneously for the drivers (D) and the passengers (P). The drivers decide which platform to work for followed by the passengers deciding which platform to use.}
    \label{fig:MPN}
\end{figure}

The terminologies used in this model are as follows:
\begin{align*}
U  &: \text{Platform U}.\\
L  &: \text{Platform L}\\
O  &: \text{Outside option (public transit)}\\
d  \in \real^+ &: \text{Average distance of a trip.}\\
\lambda \in \real^+ &: \text{Wait cost multiplier}\\
r^{u/l/o} \in \real^+ &: \text{Rate charged per mile  by }U/L/O\\
c^{u/l/o} \in \real^+ &: \text{Commission paid per mile by } U/L/O\\
p^{u/l/o} \in [0,1] &: \text{Proportions of passengers using } U/L/O\\
a^{u/l} \in [0,1] &: \text{Availability of drivers on } U/L\\
g \in \real^+ &: \text{Gas cost per mile}\\
\end{align*}
The game consists of three stages with the platforms first deciding on the rate to charge and commission to pay out per mile. Following this, the drivers collectively self-allocate themselves between the two platforms. Finally, the passengers decide which platform (or the public transit) to use depending upon the monetary and waiting costs. A node of the MPN is defined as $\{f, \xi, J\}$ where $f$ denotes the utility/cost being optimized for, $\xi$ denotes the feasible set, and $J$ denotes the endogenous variables being optimized on. The solution of a MPN depends upon both the nodes and the edges of the MPN (which establish a hierarchical relationship resembling a Stackelberg process), and we refer readers to the original work introducing MPNs for further details \cite{laine2024mathematical}. For this particular problem, the MPN nodes of figure \ref{fig:MPN} are defined as follows (maximization objectives in bold):
\newcommand{\intc}{\scalebox{1.2}{$~\cap~$}}
\begin{align}
    \label{mpn:u}
    U := \left\{ 
        \begin{aligned}
            f &: \pmb{d p^u (r^u - c^u)} \\
            \xi &:  r^u, c^u \ge 0 \\
            J &: r^u, c^u
        \end{aligned}
    \right\}
\end{align}

\begin{align}
    \label{mpn:l}
    L := \left\{ 
        \begin{aligned}
            f &: \pmb{d p^l (r^l - c^l)} \\
            \xi &: r^l, c^l \ge 0 \\
            J &: r^l, c^l
        \end{aligned}
    \right\}
\end{align}

\begin{align}
    \label{mpn:d}
    D := \left\{ 
        \begin{aligned}
            f &: \pmb{d p^u (c^u - g) + d p^l (c^l - g)} \\
            \xi &: 0 \le a^u, a^l \le 1  \intc  a^u + a^l \le p^u + p^l  \\
            J &: a^u, a^l
        \end{aligned}
    \right\}
\end{align}

\begin{align}
    \label{mpn:p}
    P := \left\{ 
        \begin{aligned}
            f &: p^u\left(dr^u + \lambda \frac{p^u}{a^u}\right) + p^l\left(dr^l + \lambda \frac{p^l}{a^l}\right) \\ &~~~+ p^o\left(dr^o + \lambda p^o\right)\\
            \xi &: 0 \le p^u, p^l, p^o \le 1 \intc p^u + p^l + p^o = 1 \\
            J &: p^u, p^l, p^o
        \end{aligned}
    \right\}
\end{align}
These nodes, in tandem with the edges shown in figure \ref{fig:MPN} define the MPN for the ridesharing problem. As can be seen, this model captures all intricacies of the target economy. The profits of platforms depend not only upon the rates they charge and the commission they pay, but also the population that use the service. But the population using this service depends upon the rates being charged and the availability of drivers, which in turn depend upon the commissions being paid out and the number of passengers willing to use the platform. Furthermore, this model also captures the network externalities. If more drivers are present, the wait cost for passengers decreases and vice-versa. Similarly, if more passengers are present, the profits for the drivers increase and vice-versa. A detailed exposition on this model can be found at \cite{koirala2023decreasing}. Below, we mention some of the key findings of this model:
\begin{enumerate}
    \item In perfectly competitive equilibrium, the solution degenerates to one where $r^u=c^u=r^l=c^l$ i.e. all platforms charge the same and pay the same. In this solution, no platform is able to earn any profit.
    \item There are two kinds of collusive outcomes possible: double-sided where platforms collude by agreeing on both rates and commission, and single-sided where platforms collude by agreeing on the bare minimum commission payout.
    \item Since the platforms obtain highest profit when commission payout is low and since it's less suspicious and more stable to maintain, single-sided collusion is the natural outcome of this game.
\end{enumerate}

\subsection{Extension to temporal multi origin-destination (OD) graphs}
Aforementioned model depicts a one-shot interaction across a single origin and destination. We extend it to a temporal multi origin-destination version suitable for use within a city-like setting. Concretely, we consider a fully-connected OD graph $G$ with $N$ nodes and $E$ edges with the corresponding edge-weights $e_{ij}$ representing the proportion of passengers at $i$ who want to move to node $j$, with $e_{ii} = 0 ~\forall i\in [N]$ where $[N]=\{1,2,..N\}$. Similarly, we also consider a distance matrix $\mathbb{D}$, similar to structure in $G$, but the edge-weights $d_{ij}$ representing the effective distance between any two nodes. These distances are allowed to be asymmetric because of traffic conditions and other extraneous factors and the distance from any node to itself is considered 0. 

\begin{figure}
    \centering
\begin{tikzpicture}[->,>=stealth,shorten >=1pt,auto,node distance=4cm,
                    thick,main node/.style={circle,draw,font=\sffamily\Large\bfseries}]

  \node[main node] (1) at (0, 0) {1};
  \node[main node] (2) at (3, 0) {2};
  \node[main node] (3) at (6, 0) {3};

  \path[every node/.style={font=\sffamily\small}]
    (1) edge node[above] {} (2)
    (2) edge node[above] {} (3)
    (3) edge[bend left=50] node[below] {$e_{31}:$ \textit{percentage of passengers in 3 who want to go to 1}} (1)
    (2) edge[bend right=50] node[below] {} (1)
    (3) edge[bend right=50] node[below] {} (2)
    (1) edge[bend left=50] node[above] {$d_{13}:$ \textit{distance from 1 to 3}} (3);
\end{tikzpicture}

    \caption{A graph with multiple origin-destination. Nodes in this graph represent a location whereas the edges indicate routes. Edge weights represent the proportions of passengers wanting to move to corresponding nodes. The distance of these routes is given by the $\mathbb{D}$ matrix.}
    \label{fig:enter-label}
\end{figure}
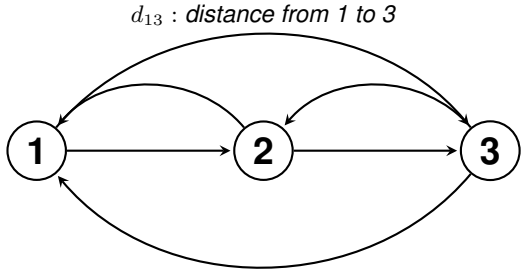

In this temporal multi OD setting, the platforms optimize on rates and commissions for each edge i.e. $r^{u/l}_{ij}(t), c^{u/l}_{ij}(t)~\forall(i,j)\in E$, the drivers optimize on their allocations for each node i.e. $a^{u/l}_i(t) ~\forall i\in[N]$, and the passengers optimize on their preference for each edge i.e. $p^{u/l/o}_{ij}(t)$ at a given time $t$. We further introduce population distribution parameters for drivers and passengers i.e. $\Pi^D, \Pi^P$ such that $\Pi^D_i(t), \Pi^P_i(t)$ indicate the \textit{number} of drivers and passengers at node $i$ at time $t$. The dynamics of these distribution parameters are defined as follows:
\newcommand{\flow}{\mathcal{F}}
\newcommand{\dflow}{\hat {\mathcal{D}}}
\newcommand{\pflow}{\hat {\mathcal{P}}}
\begin{align}
\begin{split}
    \label{eq:PIP}
    \dot \Pi^P_j(t) = &\sum_{i\in [N] \setminus \{j\}}{\left(\flow^u_{ij}(t) + \flow^l_{ij}(t) +\flow^o_{ij}(t)\right)} \\
    &-\sum_{i\in [N] \setminus \{j\}}{\left(\flow^u_{ji}(t) + \flow^l_{ji}(t) +\flow^o_{ji}(t)\right)}
\end{split}  \\
\begin{split}
    \label{eq:PID}
    \dot \Pi^D_j(t) = &\sum_{i\in [N] \setminus \{j\}}{\left(\flow^u_{ij}(t) + \flow^l_{ij}(t) \right)} \\
    &-\sum_{i\in [N] \setminus \{j\}}{\left(\flow^u_{ji}(t) + \flow^l_{ji}(t) \right)}
\end{split}  
\end{align}

Where $\flow^{u/l/o}_{ij}(t)$ indicates the instantaneous flow of traffic via the corresponding platform from node $i$ to $j$ at time $t$ defined as:
\begin{align}
    \label{eq:publicflow}
    \flow^{o}_{ij}(t) &= \pflow^o_{ij}(t) \\
    \label{eq:ulflow}
    \flow^{u/l}_{ij}(t) &= min\left(\pflow^{u/l}_{ij}(t), \dflow^{u/l}_{ij}(t)\right)
\end{align}

Where $\pflow^{u/l/o}_{ij}(t)$ and $\dflow^{u/l}_{ij}(t)$ are defined as available passenger and driver flow from node $i$ to $j$ via corresponding platforms at given time. Intuitively, equation (\ref{eq:publicflow}) encodes an assumption that outside option i.e. public transit is pervasive and all passengers willing to take public transit are accommodated at any time. Whereas, equation (\ref{eq:ulflow}) indicates that the total flow via a platform is constrained by both available flow of passengers and drivers with a one-to-one driver passenger correspondence for that platform at that time. The available flows are calculated as follows:
\begin{align}
    \pflow^{u/l/o}_{ij}(t) = \Pi^P_i(t)~e_{ij}~ {p^{u/l/o}}^*_{ij}(t) \\
    \dflow^{u/l}_{ij}(t) = \Pi^D_i(t) ~ e_{ij} ~ {a^{u/l}}^*_i(t)
\end{align}
Where ${p^{u/l/o}}^*_{ij}(t)$ and ${a^{u/l}}^*(t)$ are the optimum response of passengers and drivers at time $t$ for the given market wage and rate conditions (see subsection \ref{subsec:optimum}).

\subsection{Temporal multi OD mathematical program network}
The extension from a single-shot single OD setting to a temporal multi OD graph setting changes the underlying mathematical program network significantly. In particular, the corresponding Extended MPN (eMPN) must have multiple nodes to accommodate different OD pairs and it must also be temporal in nature. The key changes for different actors are:
\begin{enumerate}
    \item The passengers' nodes are now $N^2-N$ in number as the passengers optimize for each edge (except self-edges). The individual cost function for each node is the same as (\ref{mpn:p}) with a key difference being that the wait-time multiplier $\lambda$ differs at each node depending upon the current passenger and driver population at that node. This emphasizes the fact that passengers (say, using $U$) are more sensitive to wait-times in the event there are fewer drivers around (even if they all drive for $U$). The eMPN node $P$ then changes to:

    \begin{align}
    \label{mpn:p2}
    \hspace*{-0.7cm}P_{ij}(t) := \left\{ 
        \begin{aligned}
            f &: {p^u_{ij}(t)\left(d_{ij} r^u_{ij}(t) + \lambda_i(t) \frac{p^u_{ij}(t)}{a^u_i(t)}\right)} \\&~~~~{+ p^l_{ij}(t)\left(d_{ij}r^l_{ij}(t) + \lambda_i(t) \frac{p^l_{ij}(t)}{a^l_i(t)}\right)} \\&~~~~{+ p^o_{ij}(t)\left(d_{ij}r^o + \lambda_i(t) p^o_{ij}(t)\right)}\\
            \xi &:  0 \le p^{u/l/o}_{ij} \le 1\\
                      &\intc p^u_{ij}(t) + p^l_{ij}(t) + p^o_{ij}(t) = 1\\
                      &\intc \lambda_i(t) = \lambda \frac{\Pi^P_i(t)}{\Pi^D_i(t)} \\
            J &: \{p^u, p^l, p^o\}_{ij}(t)
        \end{aligned}
    \right\}
\end{align}

    \item There are $N$ eMPN nodes for drivers. A major difference in these new nodes is that we eschew using the \textit{matching constraint} i.e. $a^u + a^l \le p^u + p^l$ as it is already modeled by \textit{flow constraint} of equation (\ref{eq:ulflow}). It is shown in \cite{koirala2023decreasing} that drivers prefer to participate fully in the ridesharing market as long as they can break even. Therefore, we add a \textit{total-covering} constraint in this model of the form $\forall i,~ a^u_i(t) + a^l_i(t) = 1$. The eMPN drivers nodes then become (objective function is in bold because this is a maximizer).

    \begin{align}
    \label{mpn:d2}
    \hspace*{-0.7cm}D_i(t) := \left\{ 
        \begin{aligned}
            f &: \pmb{\sum_{j\in [N]}{d_{ij}p^u_{ij}(t) (c^u_{ij}(t) - g)}} \\
            &~~~~~~~~~\pmb{+d_{ij}p^l_{ij}(t) (c^l_{ij}(t) - g)}\\
            \xi &: 0 \le a^{u/l}_i(t) \le 1 \intc a^u_i(t) + a^l_i(t) = 1 \\
            J &: \{a^u, a^l\}_{i}(t)
        \end{aligned}
    \right\}
    \end{align}

    \item The platform's profit functions are now dependent upon the \textit{flows} (equations \ref{eq:ulflow}) in the graph. The relevant eMPN nodes are specified as:
    \begin{align}
    \label{mpn:ul}
    \hspace*{-1cm}\{ U/ L\}(t) := \left\{ 
        \begin{aligned}
            f &: \pmb{\sum_{(i,j)\in E}{d_{ij} \flow^{u/l}_{ij}(t) (r^{u/l}_{ij}(t) - c^{u/l}_{ij}(t))}} \\
            \xi &:  r^{u/l}_{ij}(t), c^{u/l}_{ij}(t) \ge 0 \\
            J &: \{r^{u/l}, c^{u/l}\}_{ij}
        \end{aligned}
    \right\}
    \end{align}
\end{enumerate}

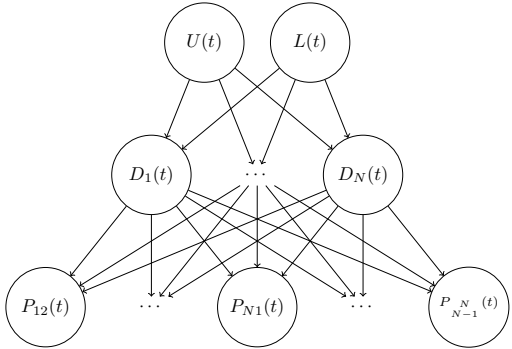
\begin{figure}[h]
    \centering
    \scalebox{.7}{\begin{tikzpicture}
        \node[minimum size=1.5cm, draw, circle] (U) at (-1,2.5) {$U(t)$};
    
        \node[minimum size=1.5cm,draw, circle] (L) at (1,2.5) {$L(t)$};
        \node[minimum size=1.5cm,draw, circle] (D1) at (-2,0) {$D_1(t)$};
        \node[draw=none] (D2) at (0, 0) {$\cdots$};
        \node[minimum size=1.5cm,draw, circle] (D3) at (2,0) {$D_N(t)$};
        
        \node[draw=none] (P1) at (-2,-2.5) {$\cdots$};
        \node[minimum size=1.5cm,draw, circle] (P2) at  (-4,-2.5) {$P_{12}(t)$};
        
        \node[draw=none] (P3) at (2,-2.5) {$\cdots$};
        
        \node[minimum size=1.5cm,draw,circle] (P4) at (0,-2.5) {$P_{N1}(t)$};
        
        \node[minimum size=1.5cm, draw, circle] (P5) at (4,-2.5) {$\scriptstyle P_{\genfrac{}{}{0pt}{}{N}{N-1}}(t)$};
        \draw[->] (U) -- (D1);
        \draw[->] (L) -- (D1);
        \draw[->] (U) -- (D2);
        \draw[->] (L) -- (D2);
        \draw[->] (U) -- (D3);
        \draw[->] (L) -- (D3);
        \draw[->] (D1) -- (P1);
        \draw[->] (D1) -- (P2);
        \draw[->] (D1) -- (P3);
        \draw[->] (D1) -- (P4);
        \draw[->] (D1) -- (P5);
        \draw[->] (D2) -- (P1);
        \draw[->] (D2) -- (P2);
        \draw[->] (D2) -- (P3);
        \draw[->] (D2) -- (P4);
        \draw[->] (D2) -- (P5);
        \draw[->] (D3) -- (P1);
        \draw[->] (D3) -- (P2);
        \draw[->] (D3) -- (P3);
        \draw[->] (D3) -- (P4);
        \draw[->] (D3) -- (P5);
        
    \end{tikzpicture}}
    \caption{eMPN for temporal multi-OD graphs. This is an extension of MPN depicted by figure \ref{fig:MPN}. There are $N$ nodes for drivers and $N^2-N$ nodes for passengers.}
    \label{fig:MPN2}
\end{figure}

The final eMPN is shown in figure \ref{fig:MPN2}.

\subsection{Optimum Response and Optimization Objective}
\label{subsec:optimum}
At each instant, after the platforms set the wages and the rates for each edge of the graph, an optimum self-allocation of the drivers and passengers takes place. For a graph with two nodes and a single edge between the two, results obtained from \cite{koirala2023decreasing} provide the theoretical optimum allocation. However, our formulation differs from \cite{koirala2023decreasing} in that the wait costs are dynamic across the edges and the driver allocation is per-node instead of per-edge. Therefore, we take the following approach to determine optimum driver and passenger allocations at a future time 
\newcommand{\dt}{\partial t}
$t+\dt$:
\begin{enumerate}
    \item The platforms choose $r^{u/l}_{ij}(t+\dt), c^{u/l}_{ij}(t+\dt)$.
    \item With, $$A^D(t) := \{a^u_{1}(t), \hdots, a^u_N(t), a^l_1(t), \hdots , a^l_N(t)\}$$ being the optimum allocation at time $t$, the drivers choose $n^D$ candidate allocations $A^D_1(t+\dt), A^D_2(t+\dt), ~ \hdots$ such that $A^D_i(t+\dt) = clip(A^D(t) + \Delta A^D_i)$. Where, the \textit{clip} function clips the elements of each array to be between 0 and 1 and $\Delta A^D_i$ is a vector of perturbations defined as:
    $$\Delta A^D_i = \{\delta a^u_1, \delta a^u_2, \hdots \delta a^u_N, 1-\delta a^u_1, \hdots 1-\delta a^u_N\}$$ Where each $\delta a^u_i \sim \mathcal{U}\left(-\delta a, +\delta a\right)$. Colloquially, $\delta a$ defines the allowed \textit{width} of the perturbation which, in turn, decides just how much the driver market may adjust in the next time-step.
    \item For each candidate allocation, the optimum passenger allocation for each edge with the per-node dynamic wait-cost multiplier $\lambda_i(t)$ is calculated according to the best response function by \cite{koirala2023decreasing} and the optimum allocations ${p^{u/l/o}}^*_{ij \in E}(t+\dt)$ are obtained.
    \item For all candidate driver allocations and their optimum passenger responses, the one that maximizes the sum of driver's profits across all nodes is chosen as the \textit{approximately optimal} driver allocation and all ${a^{u/l}}^*_{i\in[N]}(t+\dt)$ are obtained. This gradient-free approach of obtaining an approximate solution for a sequential optimization problem is in line with what's proposed in \cite{koirala2023monte}. All driver nodes are in Nash equilibrium with each other and maximizing their sum of utility i.e. finding the pareto-optimal solution, may not always yield the true Nash solution, but we still choose this method to relatively simplify the computation.
\end{enumerate}

It is evident that the hyperparameters $a_N$ and much importantly, $\delta a$, determine how the supply market (drivers) responds to exogenous changes. In particular, we take a look into the following two cases:
\subsubsection{Responsive supply market}
In this case, we set $\delta a$ to a high value (say 1) which 
causes drivers to rapidly adjust their allocation across the nodes in response to the rate/wage fluctuation of the platforms.
\subsubsection{Lagging supply market}
In this case, we set $\delta a$ to a low value (say 0.05). This causes a slow change in the drivers and can be thought of as a measure of some kind of intrinsic inertia whereby drivers want to keep on driving for platforms which they've been driving for all along. This is a reasonable stylization of the real world market. 
\begin{figure*}[htbp]
    \centering
    \begin{subfigure}{\columnwidth}
        \includegraphics[width=\linewidth]{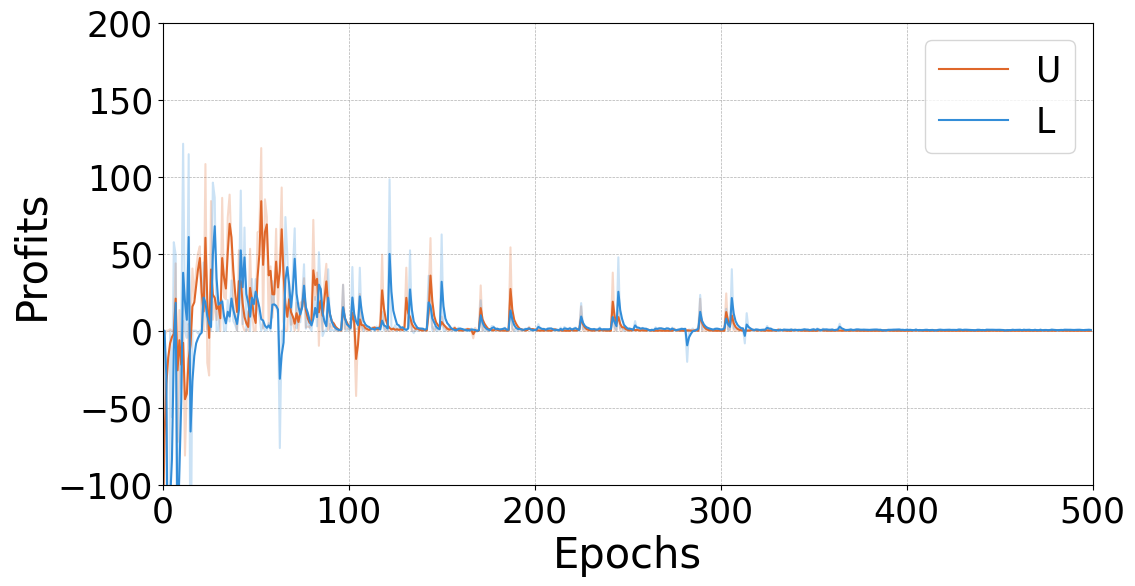}
        \caption{Profit at the end of episode for responsive market ($\delta a =1$)}
        \label{fig:sub1}
    \end{subfigure}
    \hfill  
    \begin{subfigure}{\columnwidth}
        \includegraphics[width=\linewidth]{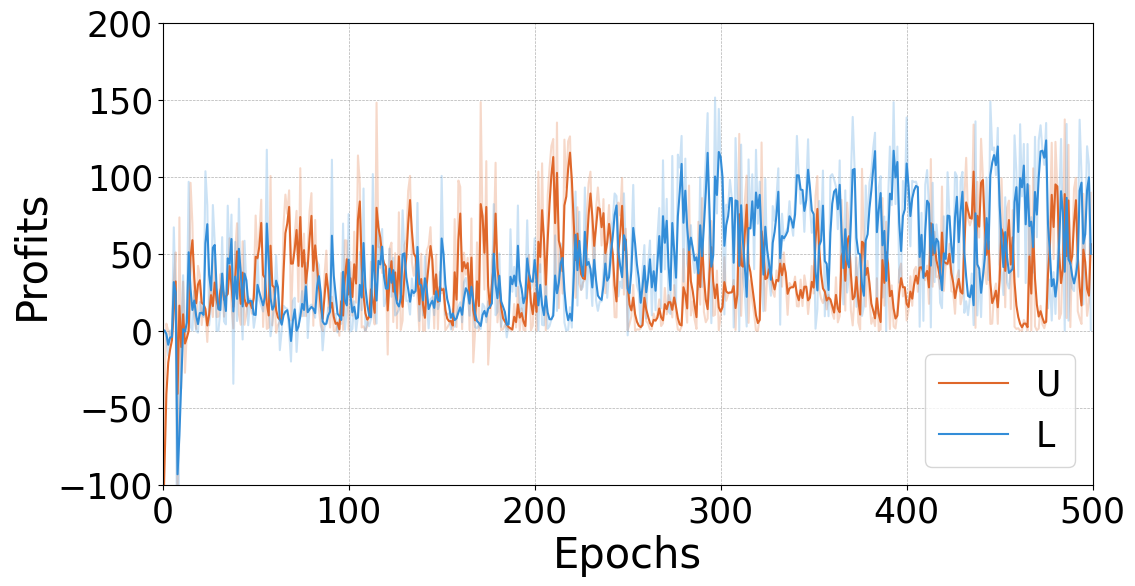}
        \caption{Profit at the end of episode for lagging market ($\delta a =0.05$)}
        \label{fig:sub2}
    \end{subfigure}
    \caption{Exponential Moving Average (EMA) smoothed instantaneous profits at the end of each episode in both cases. The profits converge as the models better learn the market response. In the responsive supply case, competition drives down the profits for both platforms but in the lagging supply case, algorithms learn to extract consistent profits by colluding.}
    \label{fig:profits}
\end{figure*}
\subsection{Optimization Objective}
In the extended rideshare model, the overarching goal of the platforms is to maximize their running profit. In other words, both platforms seek to maximize the following objective:
\begin{align}
    U^* &:= \max_{r^u_{ij}(t), c^u_{ij}(t)} \int_{t=0}^{T} {U(t) \dt} \\
    L^* &:= \max_{r^l_{ij}(t), c^l_{ij}(t)} \int_{t=0}^{T} {L(t) \dt}
\end{align}
Where $U(t), L(t)$ are as defined in equation (\ref{mpn:ul}). In practice, we discretize the model using a small time-step $\Delta t = 0.01$ for computational purposes. To solve this optimization problem, we use the undermentioned algorithm.

\subsection{Proximal Policy Optimization (PPO) for Multi-Agent Reinforcement Learning}
PPO is the state-of-the-art policy gradient based on-policy reinforcement learning algorithm that uses a surrogate objective to provide the flexibility of trust region based policy gradient methods while still being easy to implement. Due to its vast popularity, PPO has been used in domains such as robotics, trading, aviation, autonomous driving, cloud application management etc. \cite{lopes2018intelligent, funika2020automatic, bohn2019deep, ye2020automated, lin2021end}. In this paper, we refrain from discussing the exact mechanisms of the algorithm and instead, refer our readers to the seminal work that first introduced it \cite{schulman2017proximal} for additional details. We now delineate our usage of PPO for solving the discretized version of the aforementioned optimization problem.

We consider a repeated game with $t=0,1,2,...$ and so on. At each time step $t$, the market state $S_t$ is given by the distribution and the allocation of drivers and passengers across the nodes and edges of the OD graph:
\begin{align*}
    S_t =  & [\{\Pi^P\}_i, \{\Pi^D\}_i, \{a^u\}_i, \{a^l\}_i,
    \{p^u\}_{ij}, \{p^l\}_{ij}, \{p^o\}_{ij}] \\&~~~ \forall i\in[N], (i,j)\in E
\end{align*}

We implement two identical PPO based reinforcement learning algorithms $PPO^U, PPO^L$ used by, respectively, platforms $U$ and $L$ in a decentralized fashion. These algorithms do not explicitly communicate with each other. In fact, they do not even have the knowledge of the competitor's price at any moment. They merely utilize the observed market allocation state $S_t$ to produce actions $\mathcal{A}^{U/L}_t = [r^{u/l}_{ij}(t+1), c^{u/l}_{ij}(t+1)]$ in order to maximize the corresponding discounted rewards at each time step. The decisions are made simultaneously and the obtained platform-specific rates and commissions are used to induce the market response as described in subsection \ref{subsec:optimum} that, along with the dynamics in equations (\ref{eq:PIP}) and (\ref{eq:PID}), construct the state $S_{t+1}$. The process continues till defined steps i.e. episode length. At the end of each episode, the underlying policy of both agents are trained using the observed action/reward pairs. Since the transition depends upon both algorithms which are continually changing, this is a \textit{non-stationary} markov decision process, and hence, there are no theoretical guarantees of convergence.

\section{Experiment}
\label{sec:setups}
As number of OD nodes $N$ increases, the state-space becomes $3N^2+N$ and the action space for each platform becomes $2N^2-2N$. Since we are dealing with a non-stationary markov decision process, to facilitate convergence and obtain a proof of concept, we work with a small OD graph with $N=2$. The relevant adjacency matrix, the cost matrix, and the initial driver/passenger distribution is given by:
$$OD = \begin{pmatrix}
    0 & 0.9\\
    0.2 & 0
\end{pmatrix}, 
\mathbb{D} = \begin{pmatrix}
    0 & 5\\
    2 & 0
\end{pmatrix}$$
$$
\Pi^D(0) = \begin{pmatrix}
    500\\
    1000
\end{pmatrix},
\Pi^P(0) = \begin{pmatrix}
    2000\\
    3000
\end{pmatrix}
$$

We set the gas cost to $g=5$, base wait-time multiplier to $\lambda=2$, and public transit per mile cost to $r^o=10$. We also limit the rates and commissions to be between $[5, 20]$ to promote convergence, and fix the number of candidate driver allocations per step to $a_N=10$. The initial allocations are randomly initialized from the distribution ${a^u}_i, {a^l}_i \sim \mathcal{N}(0.5, 0.01)$  , $p^{u}_{ij}, p^l_{ij}, p^o_{ij} \sim \mathcal{N}(\frac{1}{3}, 0.01)$. All values are appropriately clipped to maintain relevant constraints. We then train our simulation for both responsive and lagging supply market with, respectively, $\delta a = \{1, 0.05\}$ with episode length of $2048$. At the end of each episode, the models are trained via PPO and the state is reinitialized as discussed.
\newcommand{\w}{0.8}

\clearpage
\begin{figure}[t]
    \centering
    \begin{subfigure}{\w\columnwidth}
        \includegraphics[width=\linewidth]{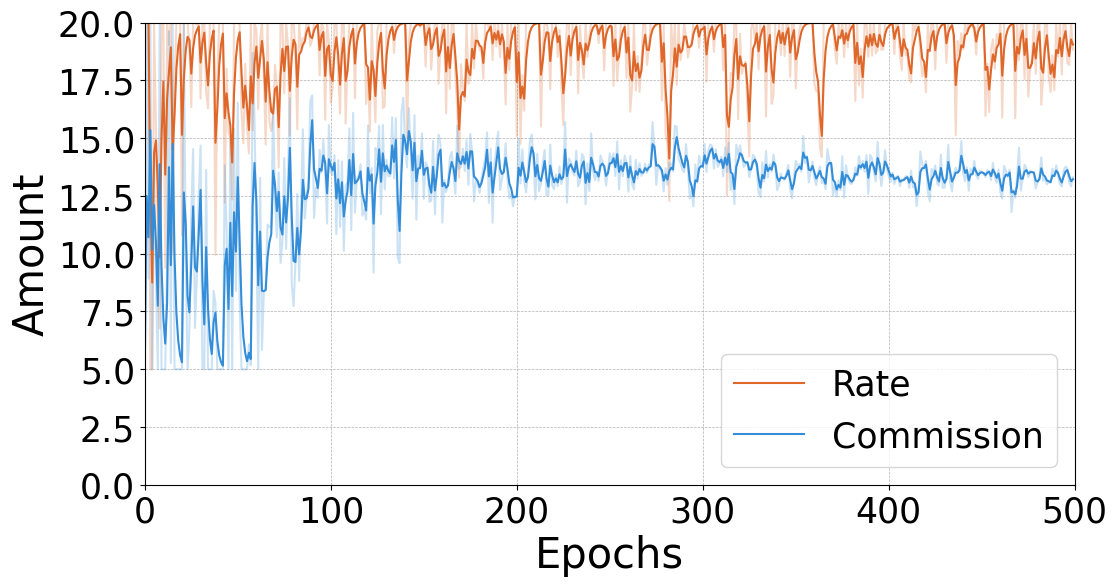}
        \caption{Rates and commissions for L on edge \textbf{0} $\rightarrow$ \textbf{1}}
        \label{fig:rlclo1responsive}
    \end{subfigure}
    \hfill
    \begin{subfigure}{\w\columnwidth}
        \includegraphics[width=\linewidth]{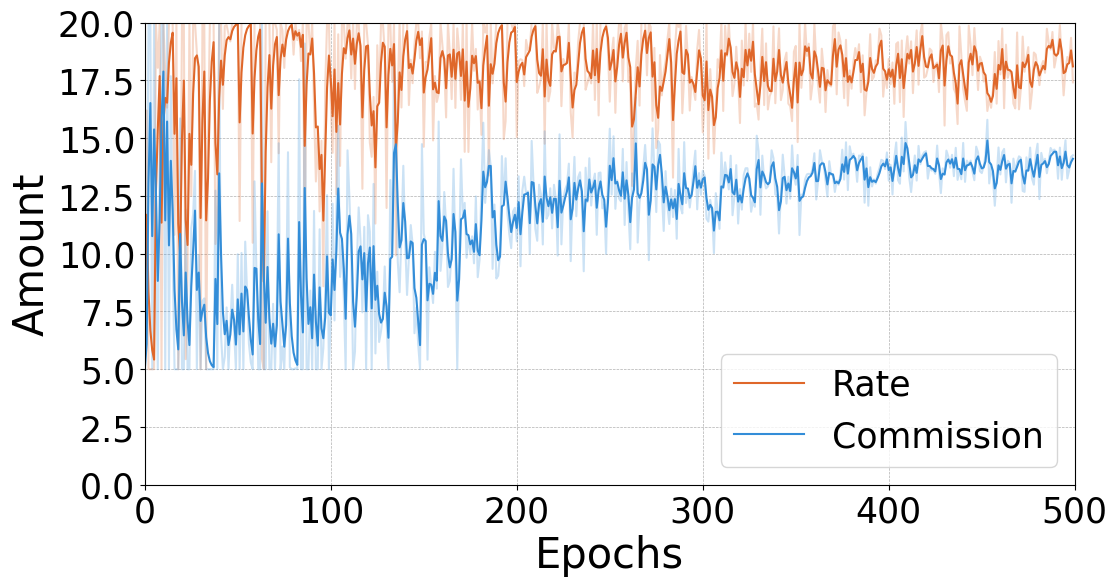}
        \caption{Rates and commissions for L on edge \textbf{1} $\rightarrow$ \textbf{0}}
        \label{fig:rlcl10responsive}
    \end{subfigure}
    \hfill
    \begin{subfigure}{\w\columnwidth}
        \includegraphics[width=\linewidth]{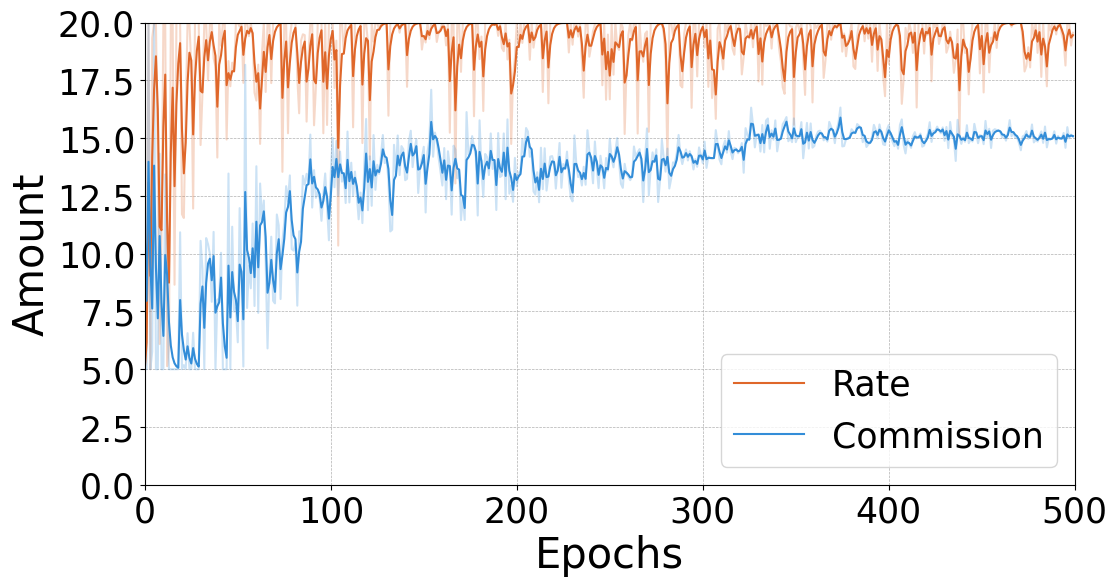}
        \caption{Rates and commissions for U on edge \textbf{0} $\rightarrow$ \textbf{1}}
        \label{fig:rucu01responsive}
    \end{subfigure}
    \hfill
    \begin{subfigure}{\w\columnwidth}
        \includegraphics[width=\linewidth]{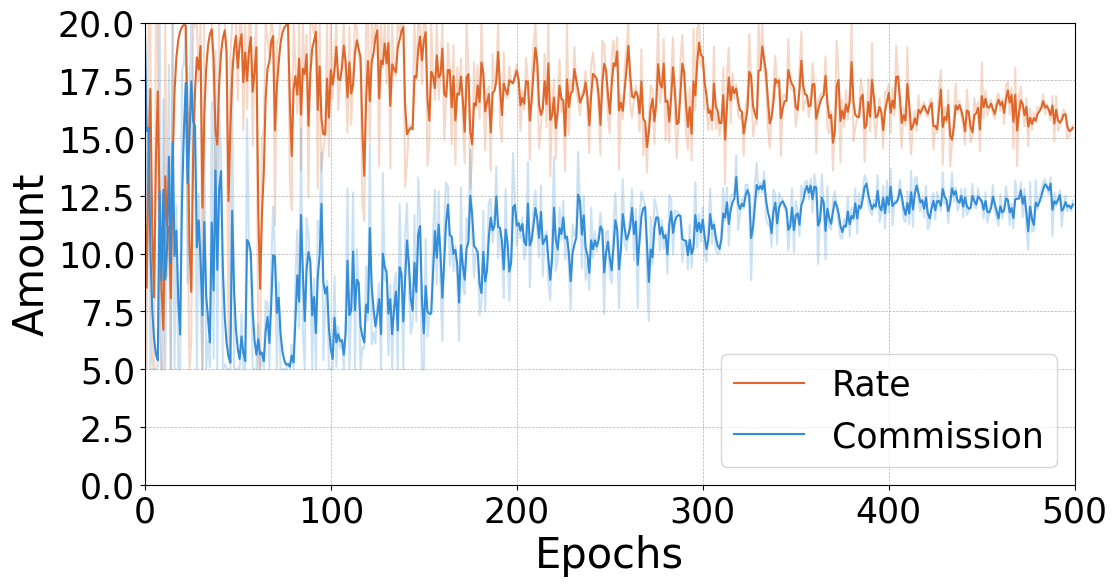}
        \caption{Rates and commissions for U on edge \textbf{1} $\rightarrow$ \textbf{0}}
        \label{fig:rucu10responsive}
    \end{subfigure}
    \caption{EMA smoothed rates and commissions for a responsive market. The competition on the supply side drives the commission towards the rate reducing platform profits.}
    \label{fig:rucu_responsive}
\end{figure}

\begin{figure}[t]
    \centering
    \begin{subfigure}{\w\columnwidth}
        \includegraphics[width=\linewidth]{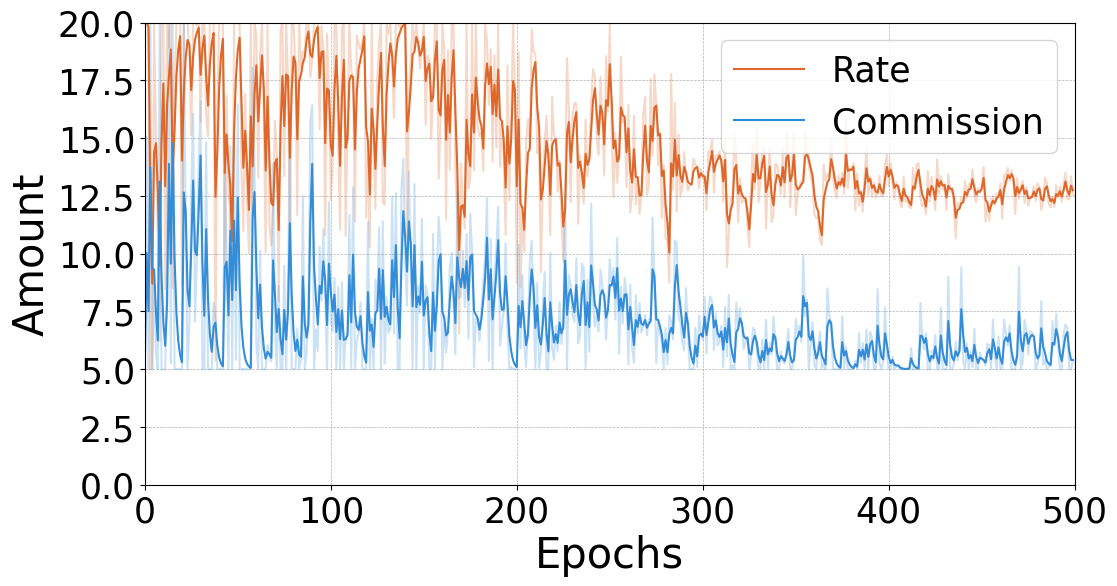}
        \caption{Rates and commissions for L on edge \textbf{0} $\rightarrow$ \textbf{1}}
        \label{fig:rlcl01lagging}
    \end{subfigure}
    \hfill
    \begin{subfigure}{\w\columnwidth}
        \includegraphics[width=\linewidth]{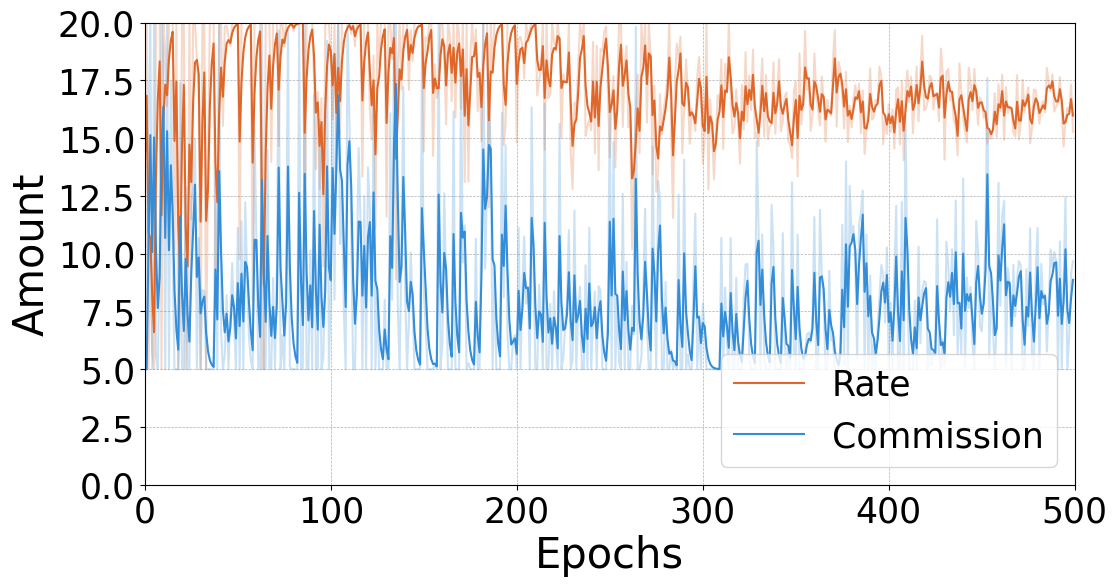}
        \caption{Rates and commissions for L on edge \textbf{1} $\rightarrow$ \textbf{0}}
        \label{fig:rlcl10lagging}
    \end{subfigure}
    \hfill
    \begin{subfigure}{\w\columnwidth}
        \includegraphics[width=\linewidth]{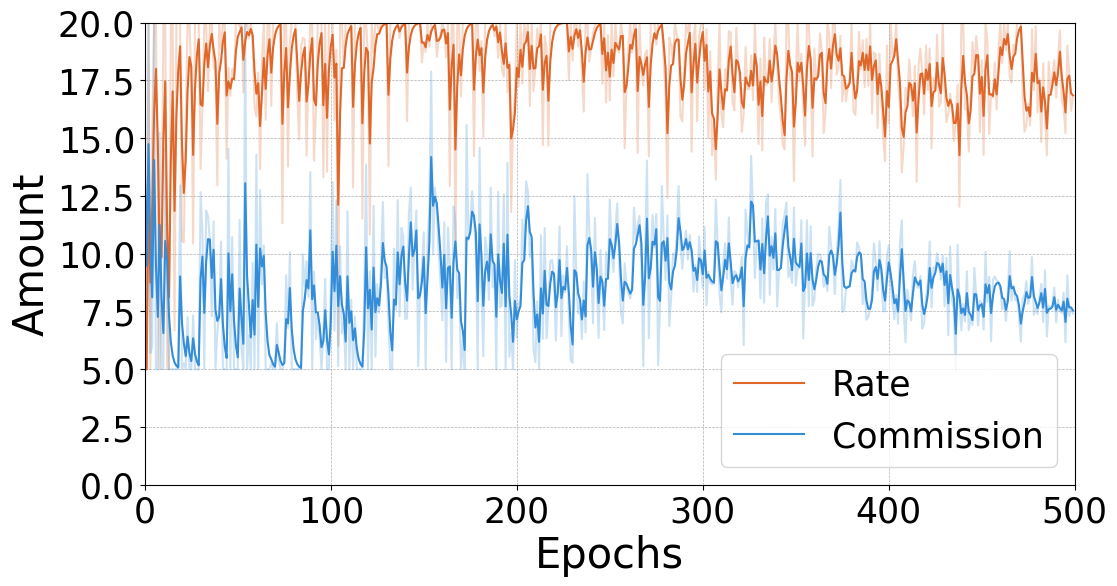}
        \caption{Rates and commissions for U on edge \textbf{0} $\rightarrow$ \textbf{1}}
        \label{fig:rucu01lagging}
    \end{subfigure}
    \hfill
    \begin{subfigure}{\w\columnwidth}
        \includegraphics[width=\linewidth]{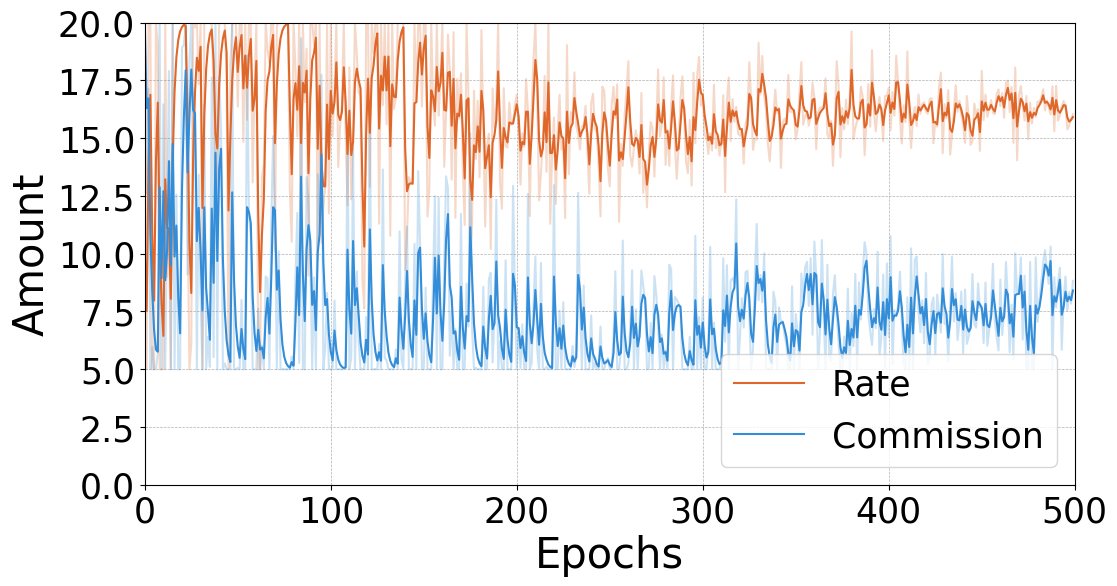}
        \caption{Rates and commissions for U on edge \textbf{1} $\rightarrow$ \textbf{0}}
        \label{fig:rucu10lagging}
    \end{subfigure}
    \caption{EMA smoothed rates and commissions for a lagging market. There are signs of collusion on the supply side (low driver commissions) leading to increased profit margins.}
    \label{fig:rucu_lagging}
\end{figure}
\noindent  We continue this process for 500 epochs. Our market model is implemented as a PettingZoo environment \cite{terry2021pettingzoo} and we base our decentralized learning algorithm on an openly available PPO implementation from \cite{seolhokim}. The code for the paper, including all config parameters for the PPO, can be found at \url{https://github.com/PraveshKoirala/RideshareMARL}

\section{Results and Discussion}
\label{sec:results}
All simulations are run on a machine with 12th Gen Intel(R) Core(TM) i7-12700 2.10 GHz processor and 32 GB of RAM. GPU was not used. The time it took for simulating both responsive and lagging market was approximately 27 minutes. The results for the simulation are depicted in figures \ref{fig:profits}, \ref{fig:rucu_responsive}, and \ref{fig:rucu_lagging}. All figures have been smoothed using Exponential Moving Average with the smoothing parameter $\alpha$ set to 0.5, in order to facilitate readability (original values are still visible). As can be observed from figure \ref{fig:profits}, the instantaneous profits for responsive market eventually dies out as compared to the profits from the lagging market towards the end of the episode. While these profits are not just governed by the rates and the commissions of the platforms but also the available passenger and driver distribution and the resulting flows (equation \ref{mpn:ul}), we theorize that the responsiveness of the drivers aids in the loss of profit as the platforms compete aggressively on the supply side by increasing the commissions in addition to the existing competition for passenger occupancy. This is substantiated by figure \ref{fig:rucu_responsive} where it can be seen that the commission payout towards the end of each episode converges towards the charged rate signifying a competition. In contrast to this, in the lagging market (fig \ref{fig:rucu_lagging}), the commissions do not significantly increase and instead, steadily converge towards the minimum driver costs ($g=5$), showing that the platforms opt to collude, rather than compete, on the supply side, earning supracompetitive profits (fig \ref{fig:profits}).

\section{Conclusion}
\label{sec:conclusions}
The obtained results have vast implications on the literature of algorithmic collusion. First and foremost, this establishes that modern algorithms like PPO have potential for emergent collusive behavior in a complex platform market. Additionally, these results also show that even when the hyperparameters for a learning algorithm are held constant, certain market characteristics like responsiveness can either induce competition or emphasize collusion. This is an interesting result for concerned regulatory bodies as this suggests that markets with high inertial (aka slow response) properties should be prioritized while investigating for potential algorithmic collusion. Lastly, this establishes that emergent collusive behaviors can negatively impact certain market segments (like drivers) more than the other. In this particular case, this result helps explain why gig workers are getting paid less despite of their indispensability in the gig economy. Future works in this area could be a large-scale simulation of this (or related) model to glean further insights into the nature of market characteristics in facilitating tacit collusion in intelligent pricing agents.

On a final note, we want to stress that this work does not accuse any real world platforms or their pricing agents of collusion. It merely provides results for a stylized model.

\addtolength{\textheight}{-12cm}   




\bibliographystyle{./IEEEtran} 
\bibliography{./IEEEexample}

\end{document}